\documentclass[apj]{emulateapj}
\slugcomment{Submitted to ApJ on March 2, 2021}
\usepackage{gensymb}
\usepackage{textcmds}
\usepackage{graphicx}
\usepackage[breaklinks=true]{hyperref}
\usepackage{epstopdf}
\usepackage{amsmath}

\DeclareGraphicsExtensions{.eps}
\shorttitle{Spin dynamics of extrasolar planets in planet-planet scattering}
\shortauthors{Hong et al.}

\begin{document}

\title{Spin dynamics of extrasolar giant planets in planet-planet scattering}
\author{Yu-Cian~Hong$^{1,2}$,Dong~Lai$^{1,2}$,Jonathan~I.~Lunine$^{1,2}$, and Philip~D.~Nicholson$^{1,2}$}
\email{ycsylva@gmail.com}

\affil{$^1$Astronomy Department, Space Sciences Building, Cornell University, Ithaca, NY 14853, USA\\
$^2$Carl Sagan Institute, Space Sciences Building, Cornell University, Ithaca, NY 14853, USA}


\vspace*{3\baselineskip}

\begin{abstract}

Planet-planet scattering best explains the eccentricity distribution of extrasolar giant planets. Past literature showed that the orbits of planets evolve due to planet-planet scattering. This work studies the spin evolution of planets in planet-planet scattering in 2-planet systems. Spin can evolve dramatically due to spin-orbit coupling made possible by the evolving spin and orbital precession during the planet-planet scattering phase. The main source of torque to planet spin is the stellar torque, and the total planet-plane torque contribution is negligible. As a consequence of the evolution of the spin, planets can end up with significant obliquity (the angle between a planet's own orbit normal and spin axis) like planets in our Solar System.

\end{abstract}

\newpage

\section{Introduction}

Planet-planet scattering is the best model to date for explaining the eccentricity distribution of extrasolar giant planets \citep{rasio,weidenschilling,lin,marzari,adams,chatterjee,ford,juric,raymond,carrera,anderson20}. In the planet-planet scattering scenario, planets are hypothesized to form in closely packed systems. Mutual perturbations trigger orbit crossings and close encounters between planets. Past literature has shown that the orbits of planets (semi-major axis, eccentricity, inclination) change instantaneously during planetary close encounters, and also evolve significantly on a short timescale outside of close encounters, before the destruction of some planets ends the scattering phase. In this paper, we show that as a consequence of planet-planet scattering, the spin of the planets evolves and planets can grow in obliquity, due to changes in the spin direction. The spin evolution is driven by interactions between a planet's own spin and orbit, which experiences changes in short time scales during planet-planet scattering. Obliquity in this work is defined as the angular separation between the spin and orbital axes of a planet. Close encounter is defined as when planets come into each other's Hill sphere.

The obliquity of a planet may provide clues to its dynamical history. Various mechanisms involving giant impacts and/or secular interatcions have been proposed to generate planetary obliquity \citep[e.g.][]{benz,dones,ward04,rogoszinski,su}. Although there is currently no direct measurement of extrasolar planetary spins and obliquities, constraints can be obtained using high-resolution spectroscopic observations \citep{snellen,bryan18,bryan20}, as well as radial velocity signals from directly imaged planets \citep{vanderburg}. High-precision photometry of transiting planets can also help constrain planetary rotations in the future \citep{seager,barnes,carter,schwartz}.

Recently, \citet{li20a} and \citet{li20b} studied giant planet spin and obliquity produced by planetary mergers. Through analytical calculation and numerical simulations, they showed that planetary collisions lead to rapidly rotating objects and a broad range of obliquities with a characteristic distribution. Two-planet scatterings typically result in planet mergers or ejection of one planet. \citet{li20a} focused on the obliquities of the former outcomes, but not the latter.  Their simulations (as well as all previous planet scattering simulations) assume the spin of a planet is constant, except when two planets merge. In this paper we show that spin-orbit coupling during the planet-planet scattering phase may 
change the spin orientation of a planet, and therefore influencing the final planetary obliquities, especially for the ejection remnants.

The next section introduces our method for simulating spin evolution in planet-planet scattering. Because orbital evolution in the planet-planet scattering phase is on timescales shorter than the secular timescale (both during and outside of planetary close encounters), spin evolution requires special treatment beyond the standard secular approaches. Section \ref{sec:result} provides a detailed view the evolution of spin and obliquity. And on the basis of four simulation sets, the section discusses statistical properties of the spin and obliquity of planets in 2-planet systems.

\section{Method}
The numerical scheme of this work builds on \citet{hong15,hong18}. The code uses \textit{Mercury} \citep{chambers} as a base, and incorporates the shape of planets and spin evolution using a non-secular equation of motion. The equation of motion for both orbit and spin evolves at every time step. Planets are simulated as oblate and symmetric around their spin axes. The code uses the built in Bulirsch-Stoer algorithm, which produces the best accuracy for close encounters. The faster version $\mbox{--}$ conservative Bulirsch-Stoer is used, since both the translational acceleration and the torque on spin depend only on position.

\subsection{Oblate planet}
The \textit{Mercury} code is modified to include planet oblateness by modifying the force term generated by the gravitational field around the planet. The gravitational field has a quadruple moment term $J_2 = 0.0147$, consistent with Jupiter and slightly less than Saturn's. All simulated planets adopt this oblateness.
The $J_2$ moment modifies the spherically symmetric force field around a point mass, to an axisymmetric force field around an extended mass. The code assumes zero reaction time of planet spin to external perturbations, but in reality there might be a time lag due to viscous dissipation for a fluid body (such as a giant planet) to react to a torque.  In general, the internal dynamics of a precessing fluid body is rather complicated, potentially involving turbulence and strong dissipation \citep{papaloizou,barker,linyufeng}. However, since the internal dissipation conserves angular momentum, we expect that as far as forced precession is concerned, the giant planet effectively behaves as a rigid body to a leading-order approximation. In any case, as in all previous works on planetary obliquities, we will ignore any complications associated with internal fluid motions in this paper.

\subsection{Spin evolution}

The spin of an oblate planet evolves under the torque from the central star and other planets. In planetary close encounters, the spin of a planet can
receive dramatically large torques in a shorter period of time than their orbital periods. Therefore, in order to treat planet-planet scattering scenarios realistically, this work adopts an instantaneous equation of motion for the spin, instead of the secular one commonly used.

For an extended mass under the gravitational influence of a point mass, different parts of it receive differing amount of gravitational attraction. The difference in gravitational attraction in different parts of the extended mass causes a torque, leading to spin evolution.  
The force on a point mass M from an extended mass m is

\begin{multline}\label{eq1}
       \vec{F} = \frac{-\,G\,M\,m}{r^2} \Biggl\{ \hat{r} - \frac{3J_{2}}{2}\,\left(\frac{R}{r}\right)^2\,[(5(\hat{r}\cdot\hat{s})^2 - 1)\,\hat{r} \\ - 2\,(\hat{r}\cdot\hat{s})\,\hat{s}] + \cdots\Biggr\},
\end{multline}
where G is the gravitational constant, r the distance between the centers of m and M, $\hat{r}$ the unit vector pointing from m to M, R the radius of m,
and $\hat{s}$ the normalized spin vector of m (eg. \citet{hilton}).

The reaction torque per moment of inertia on the extended body is 

\begin{equation}
        \frac{-\vec{r}\times\vec{F}}{I} = \frac{d}{dt}\left(\Omega\,\hat{s}\right),
\end{equation} where $\Omega$ is the rotation rate of the planet, which is set to be $\frac{2\pi}{10\ hours}$. The rotation period is close to that of Jupiter. For simplicity, the rotation rate is assumed to remain constant throughout the simulations.

Combining the two equations above, we can obtain the instantaneous equation of motion for the spin components of mass m:

\begin{equation}\label{eq3}
       \frac{d\hat{s}}{dt}= \frac{3\,G\,M}{r^3} \frac{J_{2}}{\lambda\,\Omega} \left(\hat{r}\cdot\hat{s}\right)\left(\hat{r}\times\hat{s}\right),
\end{equation}
where $\lambda = \frac{I}{mR^2}$ is the normalized moment of inertia. $\lambda$ is set to 0.25, close to that of Jupiter, in all simulations.

The sources of torque on planet spin include the central star and other planets.
The implementation of the above equation of motion into \textit{Mercury} is verified with the past evolution of Mars's obliquity. The simulation is checked against the result in \citet{ward73} using the initial conditions provided in \citet{quinn}.
The integration error limit of the simulations is set at $ 10^{-12}$ per time step, and the initial time step is 1 day. The integrator automatically adjusts the time step in accordance with the set error limit.

\section{Simulation Setting}\label{sec:setting} 

This work performs controlled planet-planet scattering experiments with 2-planet systems. Four different sets of simulations are listed in Table \ref{table1}.

\begin{table}[]
\centering
\begin{tabular}{|c|c|c|c|c|}
    \hline
    Set & Variable from base & Valid/total sims & Col. prob.\\
    \hline
    1(Base) &  Inner pl $1\ AU$ & 5486 / 18636 & 0.67 \\
    2 &  Inner pl $0.3\ AU$ & 5399 / 48430 & 0.87\\
    3 &  Inner pl $5\ AU$ & 5397 / 8538 & 0.31\\
    4 &  No pl-pl torque & 5795 / 19327 & 0.66\\
  \hline
\end{tabular}
\caption{Simulation sets}
\label{table1}
\end{table}

\begin{table}[]
\centering
\begin{tabular}{|c|c|c|}
    \hline
    Planet parameters & Inner planet & Outer planet \\
    \hline
    Semi-major axis(AU) & 1 &  1 + 2 $\times R_H$ \\
    Eccentricity & 0.002  $\mbox{--}$ 0.01  & 0.002 $\mbox{--}$ 0.01\\ 
    Inclination from z axis($^{\circ}$) & 0 & 3 \\
    Obliquity($^{\circ}$) & 0 & 0 \\
    Arg. pericenter($^{\circ}$) & 0 $\mbox{--}$ 360 & 0 $\mbox{--}$ 360 \\
    Long. of asc. node  ($^{\circ}$) & 0 $\mbox{--}$ 360 & 0 $\mbox{--}$ 360 \\
    Mean anomaly ($^{\circ}$) & 0 $\mbox{--}$ 360 & 0 $\mbox{--}$ 360 \\

  \hline\hline
  Mass & 1.5 $M_J$ & 1 $M_J$ \\
  Radius & 1 $R_J$ & 1 $R_J$\\

  \hline
\end{tabular}
\caption{Base simulation: initial conditions}
\label{table2}
\end{table}

The base simulation (Table \ref{table2}) contains a Sun-like central star, an inner planet with mass $m_1 = 1.5\ M_J$ (Jupiter mass) and an outer planet with mass $m_2 = 1\ M_J$. Their radii are equal to that of Jupiter. The inner planet is located at 1 AU, and the outer planet at a distance of 2 Hill radius from the inner planet (ie. $a_2 = a_1 + 2R_H$, with $R_H$ being the Hill radius of the inner planet). Throughout the paper, planets will be called the same name based on their initial position for consistency, although during the scattering phase an inner planet may become more distant than its counterpart. The spacing is set so that close encounters start early in the simulation. The planets are also given small initial eccentricities of $0.002\mbox{--}0.01$ to facilitate early close encounters. The inner planet's orbit normal aligns with the z-axis in the cartesian coordinate system, and the outer planet's orbit normal is inclined by $3^{\circ}$ from that of the inner planet. The two planets' spin axes are aligned with their orbit normal (ie. obliquity $= 0^{\circ}$). The planets' argument of pericenter, longitude of ascending node, and the mean anomaly are randomly distributed from $0^{\circ}$ to $360^{\circ}$. The other 3 simulation sets are constructed by varying one parameter in the base set (Table \ref{table2}). Simulation set 2 places the inner planet at 0.3 $AU$, simulation set 3 places the inner planet at 5 AU, and simulation set 4 is the same as the base except that the spin torque between planets is set to zero, but the stellar torque is retained. The outer planet in set 2$\mbox{--}$4 is also located at 2 $R_H$ away from the inner planet.  Each simulation is stopped once a planet has been removed from the system, by collision with the star or another planet, or by ejection from the system (semi-major axis $> 1000\ AU$). Both planets in the system are tracked throughout the simulation.

\section{Results}\label{sec:result} 

\subsection{Spin dynamics of planets in secular systems}
Here we will discuss spin dynamics in systems evolving under a secular timescale, and derive its application to systems evolving on shorter (non-secular) timescales in the next sub-section. In secular systems, when a planet's own spin and orbital angular momenta are under perturbation, the obliquity evolution falls into three regimes. When $\mid\frac{d\hat{s}}{dt}\mid$ (rate of change of the direction of spin momentum) is much greater than $\mid\frac{d\hat{L}}{dt}\mid$ (rate of change of the direction of orbital angular momentum), the spin axis of the planet will follow its own orbit normal, and the planet will retain a constant obliquity. If $\mid\frac{d\hat{s}}{dt}\mid$ is much smaller than $\mid\frac{d\hat{L}}{dt}\mid$, the spin of the planet doesn't follow its own orbit normal. Instead, it precesses with a constant angular separation around the normal to system's invariable plane. In 2-planet systems, the spin of a planet precesses around the total orbital angular momentum axis of the two planets. When $\mid\frac{d\hat{s}}{dt}\mid\sim\mid\frac{d\hat{L}}{dt}\mid$, resonance crossing may lead to significant spin evolution, and can generate high obliquity planets \citep[e.g.][]{storch, storch15, anderson}.  Eccentricity evolution in planet-planet scattering can also alter the rate of spin evolution, because it alters r$\mbox{--}$dependent terms in equation \ref{eq1}. Therefore, eccentricity evolution in planet-planet scattering can also contribute to resonance crossings \citep{storch}. 

\subsection{Spin dynamics in planet-planet scattering}
In planet-planet scattering, the spin of a planet evolves under torques from other bodies in the system. Because of the short time-scale evolution of planetary orbits, and very short duration of planetary close encounters, the spin evolution of the planets is governed by an instantaneous equation of motion (Eq. \ref{eq3}). 
Here we provide a detailed view on how planets can experience significant spin evolution and gain high obliquity in planet-planet scattering. During the planet-planet scattering phase, close planetary encounters, as well as strong mutual perturnation outside of planetary close encnounters, can change the planet's
orbital elements by large amounts over short periods, which in turn can change $\mid\frac{d\hat{L}}{dt}\mid$ and $\mid\frac{d\hat{s}}{dt}\mid$. This can drive the system close to a state where $\mid\frac{d\hat{L}}{dt}\mid \simeq \mid\frac{d\hat{s}}{dt}\mid$, and thus lead to large variations in the obliquity due
to stellar torques. For any inclination and spin angle presented in the result section, the reference plane is the x-y plane, which has a small tilt of about $1^{\circ}$ from the system's invariable plane.

\begin{figure}[!t]
    \centering
  \includegraphics[width=\linewidth]{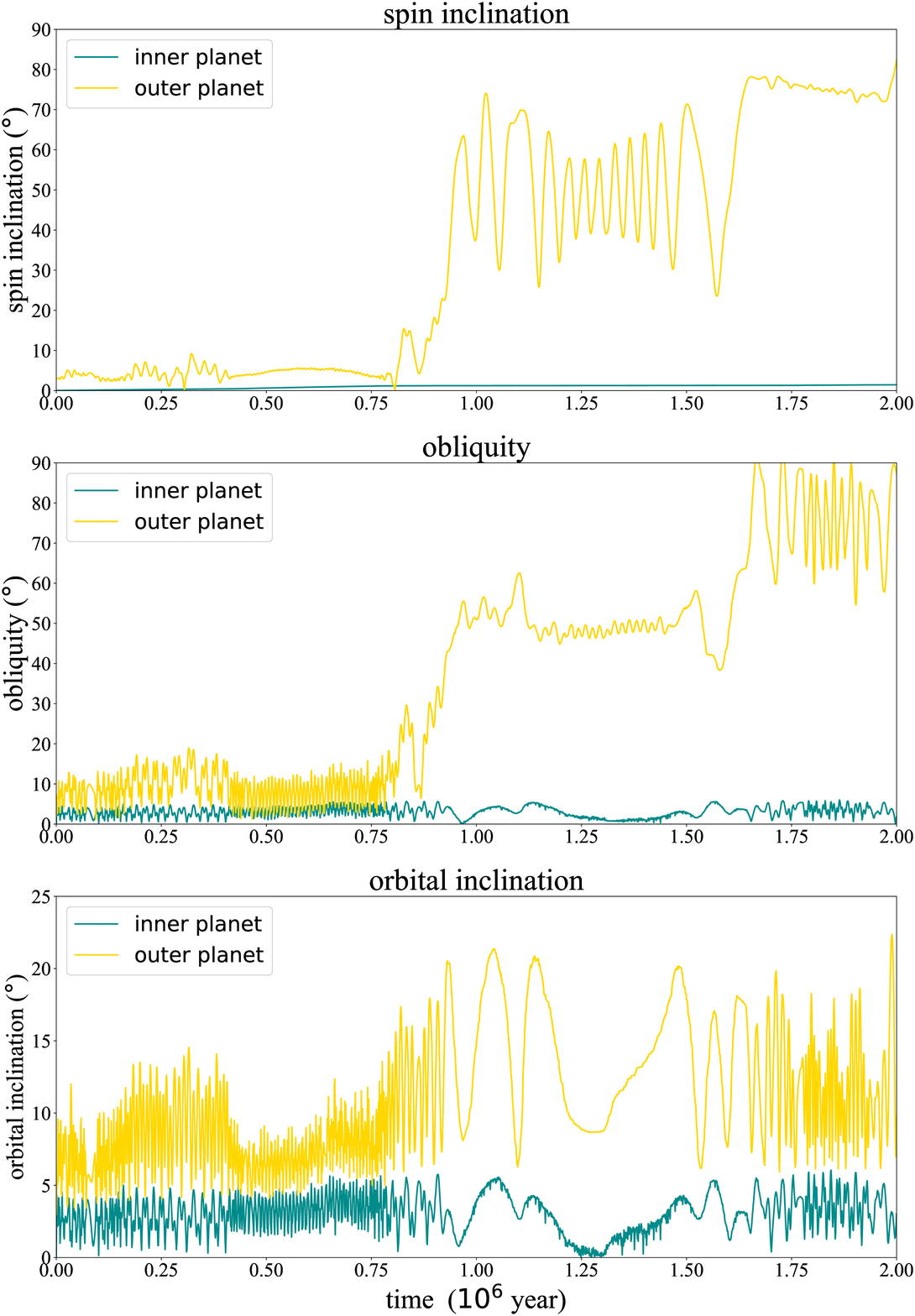}

  \caption{%
An example of the time evolution of the spin inclination (the angle between $\hat{s}$ and the z$\mbox{--}$axis), obiliquite (the angle between $\hat{s}$ and $\hat{L}$), and orbital inclination (the angle between $\hat{L}$ and the z$\mbox{--}$axis) in the base simulation. The z$\mbox{--}$axis is along the initial orital and spin axes of the inner planet, and is approximately perpendicular to the invariable plane. The naming of the inner planet and outer planet is based on their initial location. The initially inner planet becomes more distant than the initially outer planet during the scattering phase.}\label{fig1}
\end{figure}

\begin{figure}[!t]
    \centering
   \includegraphics[width=\linewidth]{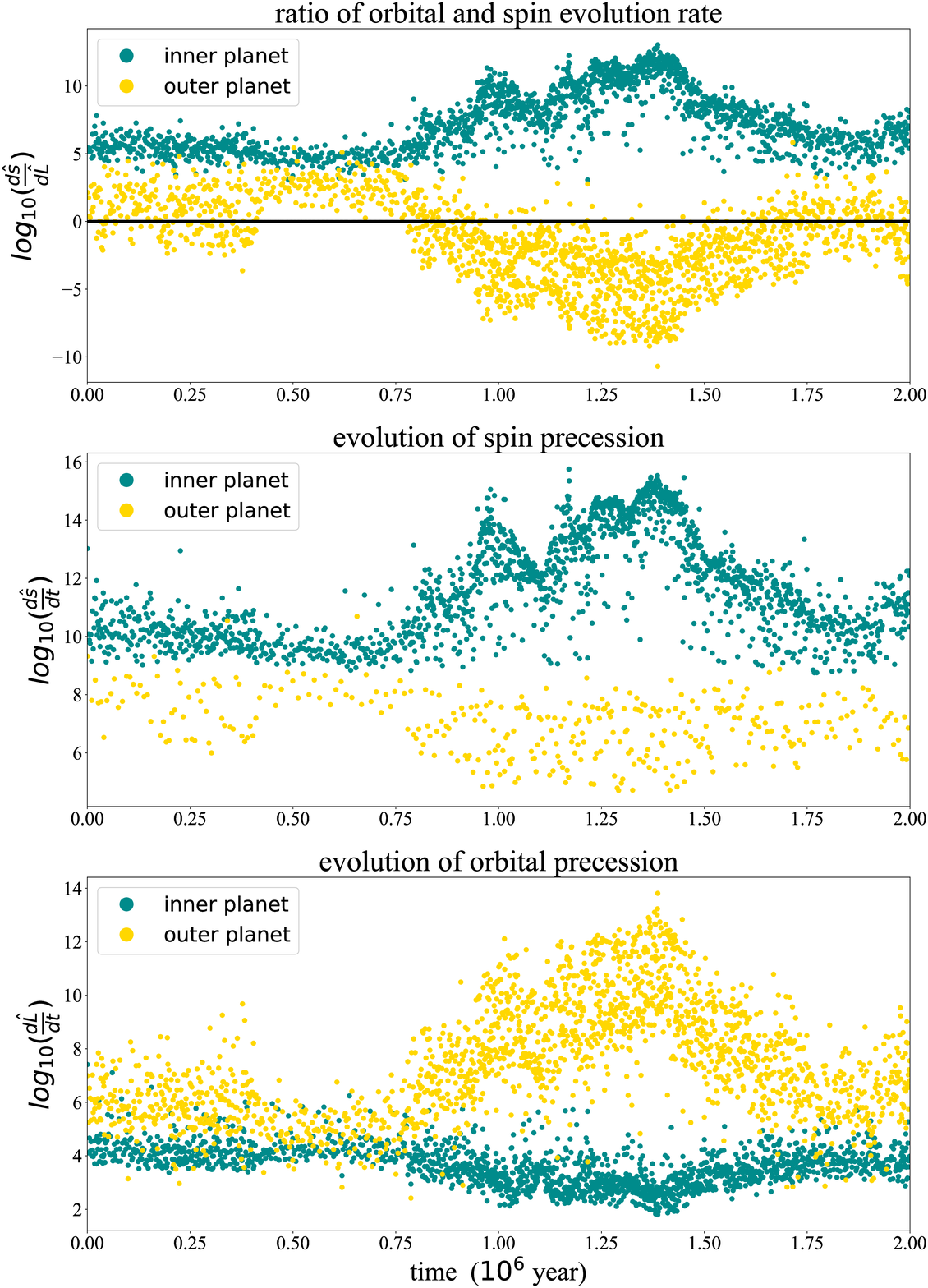}
  \caption{%
The ratio of orbital and spin evolution rate, spin evolution rate, and orbital evolution rate for the simulation depicted in Figure \ref{fig1}.
  }\label{fig4}
\end{figure}

Figure \ref{fig1} shows the spin evolution of a two$\mbox{--}$planet system in the base simulation set for the first two million years. The spin of the planets evolves chaotically. The outer planet's spin evolves beyond $80^{\circ}$ from its initial orientation at 2 million years. The inner planet's spin evolution is relatively stable. Figure \ref{fig1} also shows the corresponding evolution of inclination and obliquity. The dramatic spin evolution of the outer planet leads to its high obliquity (max $= 90^{\circ}$), under a relatively modest inclination. The inner planet experiences very little spin evolution, with a maximum displacement of $1.4^{\circ}$. The contrast in the scale of spin evolution between the outer and inner planets can be explained by considering the three regimes laid out in the last subsection. When the planet's orbital precession rate $\mid\frac{d\hat{L}}{dt}\mid$ is close to $\mid\frac{d\hat{s}}{dt}\mid$, resonance effects can cause the spin to significantly evolve. In the top plot of Figure \ref{fig4}, the ratio between  $\mid\frac{d\hat{L}}{dt}\mid$ and $\mid\frac{d\hat{s}}{dt}\mid$ for the outer planet lingers around 1 very frequently ($\rm log_{10}\mid\frac{d\hat{L}}{d\hat{s}}\mid \sim0$), so its spin evolves significantly. $\mid\frac{d\hat{L}}{dt}\mid$ and $\mid\frac{d\hat{s}}{dt}\mid$ have chances to be close to each other because in planet-planet scattering, both can evolve significantly, as shown in the middle and bottom plot of Figure \ref{fig4}. 

For the inner planet, $\mid\frac{d\hat{L}}{d\hat{s}}\mid$ stays much higher than 1 for the most part, so its spin evolves very little. Section \ref{subsec:statistics} will show that the main contributor for $\mid\frac{d\hat{s}}{dt}\mid$ is the central star.

\subsection{Statistics}\label{subsec:statistics}

\begin{figure}[!t]
    \centering
   \includegraphics[width=\linewidth]{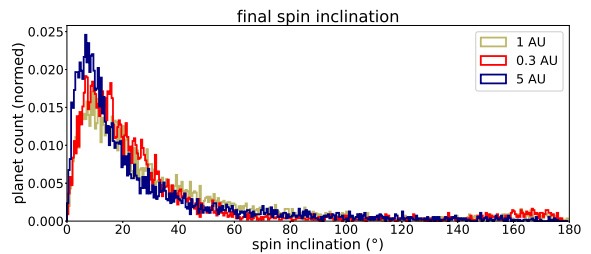}
  \caption{%
 The distribution of final spin inclination of the remnant planet after the other planet is ejected, for simulation set 1$\mbox{--}$3, where the initial semimajor axis of the inner planet is at 1, 0.3 and 5 $AU$, respectively. The result is obtained from $\sim5500$ simulations in each set that experience the ejection of one planet and retains a surviving planet. The total number of runs is shown in Table \ref{table1}.
  }\label{s}
\end{figure}

\begin{figure}[!t]
    \centering
   \includegraphics[width=\linewidth]{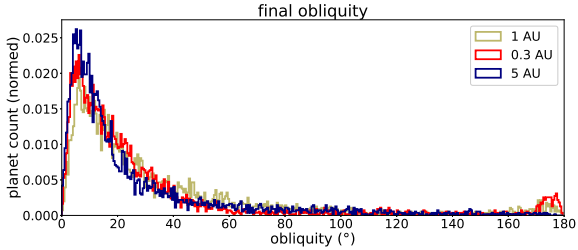}
  \caption{%
  Same as Fig.\ref{s}, but for the distribution of the final planet obliquity.
  }
  \label{o}
\end{figure}

\begin{figure}[!t]
    \centering
   \includegraphics[width=\linewidth]{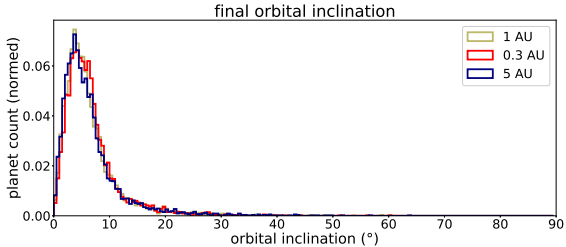}
  \caption{%
  Same as Fig.\ref{s}, but for the distribution of final orbital inclination. The K$\mbox{--}$S statistic
  of 0.06 shows that the distributions of simlation set 1 and 2 (1 $AU$ and 0.3 $AU$) have close resemblance with a p$\mbox{--}$value of 0.94.
  }
  \label{i}
\end{figure}

This section discusses the results of planet-planet scattering in systems with two giant planets in 4 simulation sets (Table \ref{table1}). The statistics use only simulations with one surviving planet after the other planet is ejected via planet-planet scattering ($23.26\%$ of all simulations)(see Table \ref{table1}), which is around 5500 simulations in each set. Simulations with planet$\mbox{--}$star ($2.71\%$ of all simulations) or planet$\mbox{--}$planet collisions ($73.61\%$ of all simulations) (see Table \ref{table1}) are excluded from the statistics, as obliquity evolution for the collision of fluid bodies is beyond the scope of this work (see \citet{li20a,li20b}). Simulations with energy error $\mid\frac{\delta E}{E}\mid$ greater than $\mathcal{O}(10^{-4})$ are also excluded ($0.36\%$ of all simulations), which is an adequate threshold for multi-planet systems \citep{barnes04, raymond}. A negligible fraction of systems ($0.03\%$ of all simulations) that don't experience planet destruction in $10^8$ years are also excluded. Figures \ref{s} and \ref{o} show the distribution of the surviving planet's final spin orientation and obliquity for simulation set 1$\mbox{--}$3. The spin angle is measured relative to the z axis, which is aligned with the inner planet's initial spin and orbit normal. The final spin and obliquity distribution is quite broad compared with the orbital inclination distribution (Fig. \ref{i}). Overall, the final spin and obliquity distributions are broader than the inclination distributions , with $19.51\%$ planets' obliquity beyond $40^{\circ}$, and $8.77\%$ of the surviving planets are on retrograde obliquity (i.e. the rotation is in reversed direction relative to the orbit). The inclination distribution only has $0.12\%$ planets beyond $40^{\circ}$. The rate of highly oblique planets $(> 40^{\circ})$ are $17.86\%,\ 31.98\%,\ 10.43\%$ in set 1, 2, and 3. There are $5.07\%$ surviving planets on retrograde obliquity in the base set, $21.46\%$ in set 2, and $3.21\%$ in set 3. The distribution of obliquity for set 3 is more clustered at lower regions compared to the other sets. From the rate of high and retrograde obliquity in set 2 (inner planet ends up at 0.19 $AU$ after the outer planet is ejected), we may expect a high rate of highly oblique planets close to the central star, and some oblique planets farther out (Figs. \ref{s} $\&$ \ref{o}). A moderate peak in Figure \ref{o} for close-in planets in set 1 and 2 between obliquities $170\mbox{--}180^{\circ}$ implies that Venus-like obliquity is possible. The simulation results imply that planets closer to the star receive more torque from it and become more evolved in obliquity.

\begin{figure}[!t]
    \centering
   \includegraphics[width=\linewidth]{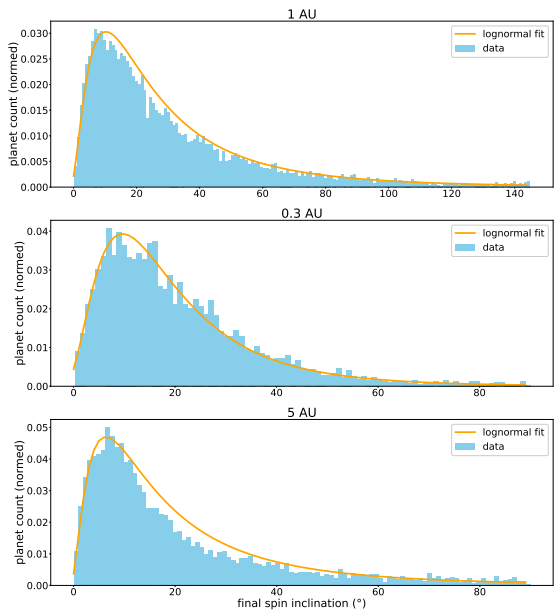}
  \caption{%
Lognormal fit of final spin inclination of set 1$\mbox{--}$4. Set 4 is mixed into the base set in the 1 $AU$ plot, as their distributions are statistically identical. The fit parameters are given in Table \ref{table3}.
  }
   \label{lognorm}
\end{figure}

The final spin distributions of the 3 simulation sets all fit best with a log$\mbox{--}$normal distribution (Fig. \ref{lognorm}):

\begin{equation}
        f (s, \sigma) = \frac{1}{\sqrt{2\pi}\sigma} e^{-\frac{1}{2} \left( \frac{ln(s)-\mu}{\sigma} \right)^2},
\end{equation}

$\sigma$ is the standard deviation, and $\mu$ the mean of the lognormal distribution.
Table \ref{table3} shows the statistics of the log-normal fit.

\begin{table}[]
\centering
\begin{tabular}{|c|c|c|c|c|}
    \hline
    Set & $\mu$ & std-dev & KS-statistics & p-value\\
    \hline
    base (1 AU) & 22.55 & 0.84 & 0.047 & 0.99  \\
    2 (0.3 AU) & 19.71 & 0.63 & 0.079 & 0.95 \\ 
    3 (5 AU) & 14.40 & 0.90 & 0.085 & 0.84 \\ 
  \hline
\end{tabular}
\caption{Log-normal fit of final spin inclination}
\label{table3}
\end{table}

\begin{figure}[!t]
    \centering
   \includegraphics[width=\linewidth]{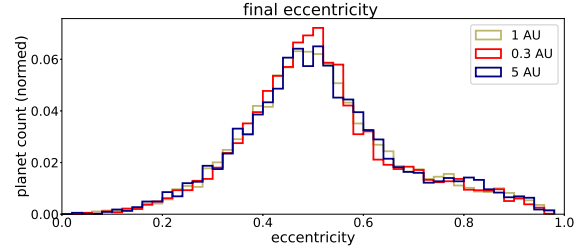}
  \caption{%
  Same as Fig.\ref{s}, but for the distribution of the final eccentricity of planets.
  }\label{ef_all}
\end{figure}

For completeness, the K-S test shows that the final eccentricity distributions between the three simulation sets are similar, with K$\mbox{--}$S statistics 0.08$\mbox{--}$0.10 and p$\mbox{--}$values 0.96$\mbox{--}$1.0 (Fig. \ref{ef_all}). The low value of K-S statistics implies the distributions are similar, and the p-value is above our set $95\%$ confidence level, so one cannot reject the null hypothesis that the distributions are the same. This result agrees with previous findings \citep{li20b}. For the total number of close encounters, close-in planets in set 2 experience more than the other 2 sets, and set 1 and set 3 has nearly identical distributions as shown by K-S test.

\begin{figure}[!t]
    \centering
   \includegraphics[width=\linewidth]{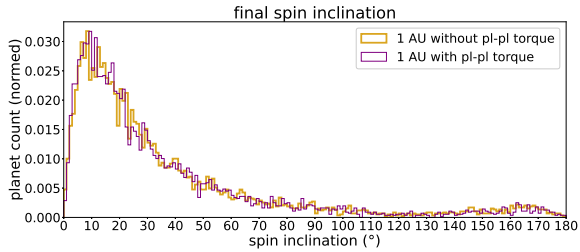}
  \caption{%
The distributions of final spin inclination of base set and set 4. The K$\mbox{--}$S statistic between the two distributions is 0.023 with a p$\mbox{--}$value of 1.0.
  }\label{set1vs4}
\end{figure}

\begin{figure}[!t]
    \centering
   \includegraphics[width=\linewidth]{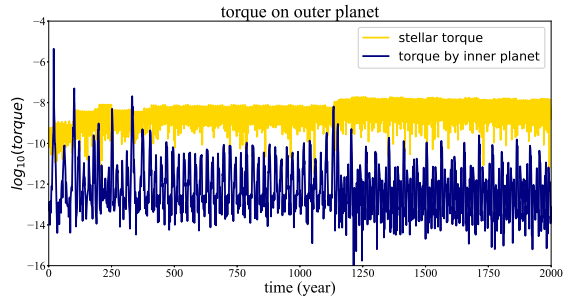}
  \caption{%
High resolution plot for the first 2000 years of evolution of torques on the outer planet from the star and the inner planet, for the simulation depicted in Figure \ref{fig1}.
  }\label{torq}
\end{figure}

Simulation set 4 has the same initial setting as the base set, but torques between planets are turned off. Figure \ref{set1vs4} shows the histograms of final spin inclinations for the base set and set 4. The distributions have a K-S statistic of 0.023 and p$\mbox{--}$value 1.00. The statistics can't reject the null hypothesis that the two distributions are the same, and the statistical evidence is very strong that planet-planet torque is very insignificant. As shown in Fig. \ref{torq}, we can  see that the stellar torque is greater than planet-planet torque most of the time. Although the torque between planets is very large during close encounters (At the first blue peak $\sim20 yr$, the torque from the inner planet is $\sim10^5$ greater than the stellar torque), due to their short duration and small mass of planets relative to the star, statistically the summed effect of planet-planet torque isn't very important. In the simulation shown in Fig. \ref{torq}, the summed torque on the outer planet from the star is $3 \times 10^4$ greater than from the inner planet in the first 2 million years of simulation time, so planet-planet torque is rather insignificant compared to stellar torque. The results show that planet-planet torque doesn't statistically affect simulation results with regards to spin or obliquity, and helps conclude that the torque from the central star is the main driver for the spin and obliquity evolution of the planets.

\section{Conclusion}\label{sec:conclusion} 
We have carried out numerical simulations of scatterings between two giant planets, keeping track of the spin evolution of each planet due to non-secular spin-orbit coupling. Our results show that planet-planet scattering is a viable mechanism for tilting planets, and does provide some probability of producing planets with retrograde rotations, especially for planets that are close-in during the scattering phase. Spin-orbit coupling can be triggered by the orbital evolution during the planet-planet scattering phase, and leads to chaotic evolution of spin inclination and obliquity. The final spin distribution of the remaining planet in an initially two-planet system fits best with log-normal distributions. The main driver for the spin evolution of planets is the central star, accompanied by eccentricity evolution from planet-planet scattering. Although planet-planet purturbations are strong during planet-planet scattering, due to reasons discussed in the last section, planet-planet perturbation has little contribution to the spin evolution. Planetary collision is outside the scope of this work (see \citet{li20a,li20b}), but it may be capable of producing highly oblique planets. The obliquity evolution of Solar System planets in planet-planet scattering may provide interesting insights on the history of our Solar System.

\section{acknowledgements}
The work was supported by the JWST Project through grant
NNX17AL71A. DL is supported in part by NASA grant 80NSSC19K0444.

\vspace*{3\baselineskip}

\bibliographystyle{apj}
\bibliography{p3_no_eprint}

\begin{thebibliography}{}
\expandafter\ifx\csname natexlab\endcsname\relax\def\natexlab#1{#1}\fi

\bibitem[{Adams \& Laughlin(2003)}]{adams}
Adams, F.~C., \& Laughlin, G. 2003, Icarus, 163, 290

\bibitem[{{Anderson} {et~al.}(2020){Anderson}, {Lai}, \& {Pu}}]{anderson20}
{Anderson}, K.~R., {Lai}, D., \& {Pu}, B. 2020, Monthly Notices of the Royal
  Astronomical Society, 491, 1369

\bibitem[{{Anderson} {et~al.}(2016){Anderson}, {Storch}, \& {Lai}}]{anderson}
{Anderson}, K.~R., {Storch}, N.~I., \& {Lai}, D. 2016, Monthly Notices of the
  Royal Astronomical Society, 456, 3671

\bibitem[{Barker(2016)}]{barker}
Barker, A.~J. 2016, Monthly Notices of the Royal Astronomical Society, 460,
  2339

\bibitem[{Barnes \& Fortney(2003)}]{barnes}
Barnes, J.~W., \& Fortney, J.~J. 2003, The Astrophysical Journal, 588, 545

\bibitem[{Barnes \& Quinn(2004)}]{barnes04}
Barnes, R., \& Quinn, T. 2004, The Astrophysical Journal, 611, 494

\bibitem[{{Benz} {et~al.}(1989){Benz}, {Slattery}, \& {Cameron}}]{benz}
{Benz}, W., {Slattery}, W.~L., \& {Cameron}, A.~G.~W. 1989, Meteoritics, 24,
  251

\bibitem[{Bryan {et~al.}(2018)Bryan, Benneke, Knutson, Batygin, \&
  Bowler}]{bryan18}
Bryan, M.~L., Benneke, B., Knutson, H.~A., Batygin, K., \& Bowler, B.~P. 2018,
  Nature Astronomy, 2, 138

\bibitem[{Bryan {et~al.}(2020)Bryan, Chiang, Bowler, Morley, Millholland,
  Blunt, Ashok, Nielsen, Ngo, Mawet, \& Knutson}]{bryan20}
Bryan, M.~L., Chiang, E., Bowler, B.~P., {et~al.} 2020, The Astronomical
  Journal, 159, 181

\bibitem[{Carrera {et~al.}(2019)Carrera, Raymond, \& Davies}]{carrera}
Carrera, D., Raymond, S.~N., \& Davies, M.~B. 2019, Astronomy \& Astrophysics,
  629, L7

\bibitem[{Carter \& Winn(2010)}]{carter}
Carter, J.~A., \& Winn, J.~N. 2010, The Astrophysical Journal, 716, 850

\bibitem[{{Chambers}(1999)}]{chambers}
{Chambers}, J.~E. 1999, Monthly Notices of the Royal Astronomical Society, 304,
  793

\bibitem[{{Chatterjee} {et~al.}(2008){Chatterjee}, {Ford}, {Matsumura}, \&
  {Rasio}}]{chatterjee}
{Chatterjee}, S., {Ford}, E.~B., {Matsumura}, S., \& {Rasio}, F.~A. 2008, The
  Astrophysical Journal, 686, 580

\bibitem[{Dones \& Tremaine(1993)}]{dones}
Dones, L., \& Tremaine, S. 1993, Icarus, 103, 67

\bibitem[{Ford \& Rasio(2008)}]{ford}
Ford, E.~B., \& Rasio, F.~A. 2008, The Astrophysical Journal, 686, 621

\bibitem[{Hilton(1991)}]{hilton}
Hilton, J.~L. 1991, The Astronomical Journal, 102, 1510

\bibitem[{Hong {et~al.}(2018)Hong, Raymond, Nicholson, \& Lunine}]{hong18}
Hong, Y.-C., Raymond, S.~N., Nicholson, P.~D., \& Lunine, J.~I. 2018, The
  Astrophysical Journal, 852, 85

\bibitem[{Hong {et~al.}(2015)Hong, Tiscareno, Nicholson, \& Lunine}]{hong15}
Hong, Y.-C., Tiscareno, M.~S., Nicholson, P.~D., \& Lunine, J.~I. 2015, Monthly
  Notices of the Royal Astronomical Society, 449, 828

\bibitem[{{Juri{\'c}} \& {Tremaine}(2008)}]{juric}
{Juri{\'c}}, M., \& {Tremaine}, S. 2008, The Astrophysical Journal, 686, 603

\bibitem[{Li \& Lai(2020)}]{li20a}
Li, J., \& Lai, D. 2020, The Astrophysical Journal, 898, L20

\bibitem[{Li {et~al.}(2020)Li, Lai, Anderson, \& Pu}]{li20b}
Li, J., Lai, D., Anderson, K.~R., \& Pu, B. 2020, Monthly Notices of the Royal
  Astronomical Society, 501, 1621

\bibitem[{Lin \& Ida(1997)}]{lin}
Lin, D. N.~C., \& Ida, S. 1997, The Astrophysical Journal, 477, 781

\bibitem[{Lin \& Ogilvie(2017)}]{linyufeng}
Lin, Y., \& Ogilvie, G.~I. 2017, Monthly Notices of the Royal Astronomical
  Society, 468, 1387

\bibitem[{{Marzari} \& {Weidenschilling}(2002)}]{marzari}
{Marzari}, F., \& {Weidenschilling}, S.~J. 2002, \icarus, 156, 570

\bibitem[{Papaloizou \& Pringle(1982)}]{papaloizou}
Papaloizou, J. C.~B., \& Pringle, J.~E. 1982, Monthly Notices of the Royal
  Astronomical Society, 200, 49

\bibitem[{Quinn {et~al.}(1991)Quinn, Tremaine, \& Duncan}]{quinn}
Quinn, T.~R., Tremaine, S., \& Duncan, M. 1991, The Astronomical Journal, 101,
  2287

\bibitem[{Rasio \& Ford(1996)}]{rasio}
Rasio, F.~A., \& Ford, E.~B. 1996, Science, 274, 954

\bibitem[{{Raymond} {et~al.}(2010){Raymond}, {Armitage}, \&
  {Gorelick}}]{raymond}
{Raymond}, S.~N., {Armitage}, P.~J., \& {Gorelick}, N. 2010, The Astrophysical
  Journal, 711, 772

\bibitem[{Rogoszinski \& Hamilton(2020)}]{rogoszinski}
Rogoszinski, Z., \& Hamilton, D.~P. 2020, The Astrophysical Journal, 888, 60

\bibitem[{Schwartz {et~al.}(2016)Schwartz, Sekowski, Haggard, Pallé, \&
  Cowan}]{schwartz}
Schwartz, J.~C., Sekowski, C., Haggard, H.~M., Pallé, E., \& Cowan, N.~B.
  2016, Monthly Notices of the Royal Astronomical Society, 457, 926

\bibitem[{Seager \& Hui(2002)}]{seager}
Seager, S., \& Hui, L. 2002, The Astrophysical Journal, 574, 1004

\bibitem[{Snellen {et~al.}(2014)Snellen, Brandl, de~Kok, Brogi, Birkby, \&
  Schwarz}]{snellen}
Snellen, I. A.~G., Brandl, B.~R., de~Kok, R.~J., {et~al.} 2014, Nature, 509, 63

\bibitem[{{Storch} {et~al.}(2014){Storch}, {Anderson}, \& {Lai}}]{storch}
{Storch}, N.~I., {Anderson}, K.~R., \& {Lai}, D. 2014, Science, 345, 1317

\bibitem[{{Storch} \& {Lai}(2015)}]{storch15}
{Storch}, N.~I., \& {Lai}, D. 2015, Monthly Notices of the Royal Astronomical
  Society, 448, 1821

\bibitem[{Su \& Lai(2020)}]{su}
Su, Y., \& Lai, D. 2020, The Astrophysical Journal, 903, 7

\bibitem[{Vanderburg {et~al.}(2018)Vanderburg, Rappaport, \& Mayo}]{vanderburg}
Vanderburg, A., Rappaport, S.~A., \& Mayo, A.~W. 2018, The Astronomical
  Journal, 156, 184

\bibitem[{Ward(1973)}]{ward73}
Ward, W.~R. 1973, Science, 181, 260

\bibitem[{Ward \& Hamilton(2004)}]{ward04}
Ward, W.~R., \& Hamilton, D.~P. 2004, The Astronomical Journal, 128, 2501

\bibitem[{Weidenschilling \& Marzari(1996)}]{weidenschilling}
Weidenschilling, S.~J., \& Marzari, F. 1996, Nature, 384, 619

\end{thebibliography}

\end{document}